\documentclass[sigconf]{acmart}
\AtBeginDocument{%
  \providecommand\BibTeX{{%
    \normalfont B\kern-0.5em{\scshape i\kern-0.25em b}\kern-0.8em\TeX}}}

\setcopyright{acmlicensed}
\copyrightyear{2018}
\acmYear{2018}
\acmDOI{XXXXXXX.XXXXXXX}

\acmConference[FAccT '24]{Make sure to enter the correct
  conference title from your rights confirmation emai}{June 03--06,
  2024}{Rio de Janeiro, Brazil}
\acmBooktitle{FAccT '24: ACM Conference on Fairness, Accountability, and Transparency (ACM FAccT), June 03--06, 2024, Rio de Janeiro, Brazil} 
\acmISBN{979-8-4007-0192-4}

\usepackage{enumitem} 
\usepackage{caption} 

\usepackage{xspace}
\newcommand{\tool}{MiMICRI\xspace}

\copyrightyear{2024}
\acmYear{2024}
\setcopyright{rightsretained}
\acmConference[FAccT '24]{The 2024 ACM Conference on Fairness, Accountability, and Transparency}{June 3--6, 2024}{Rio de Janeiro, Brazil}
\acmBooktitle{The 2024 ACM Conference on Fairness, Accountability, and Transparency (FAccT '24), June 3--6, 2024, Rio de Janeiro, Brazil}\acmDOI{10.1145/3630106.3659011}
\acmISBN{979-8-4007-0450-5/24/06}

\begin{document}

\title[\tool: Towards Domain-centered Counterfactual Explanations of Cardiovascular Image Classification Models]{\tool: Towards Domain-centered Counterfactual Explanations of Cardiovascular Image Classification Models}

\author{Grace Guo}
\orcid{0000-0001-8733-6268}
\affiliation{%
  \institution{Georgia Institute of Technology}
  \streetaddress{North Ave NW}
  \city{Atlanta}
  \state{Georgia}
  \country{USA}
  \postcode{30332}
}
\email{gguo31@gatech.edu}

\author{Lifu Deng}
\orcid{0000-0003-2475-8396}
\affiliation{%
  \institution{Cleveland Clinic}
  \streetaddress{9500 Euclid Avenue}
  \city{Cleveland}
  \state{Ohio}
  \country{USA}
  \postcode{44195}
}
\email{dengl2@ccf.org}

\author{Animesh Tandon}
\orcid{0000-0001-9769-8801}
\affiliation{%
  \institution{Cleveland Clinic}
  \streetaddress{9500 Euclid Avenue}
  \city{Cleveland}
  \state{Ohio}
  \country{USA}
  \postcode{44106}
}
\email{tandona2@ccf.org}

\author{Alex Endert}
\orcid{0000-0002-6914-610X}
\affiliation{%
  \institution{Georgia Institute of Technology}
  \streetaddress{North Ave NW}
  \city{Atlanta}
  \state{Georgia}
  \country{USA}
  \postcode{30332}
}
\email{endert@gatech.edu}

\author{Bum Chul Kwon}
\orcid{0000-0002-9391-6274}
\affiliation{%
  \institution{IBM Research}
  \streetaddress{314 Main St.}
  \city{Cambridge}
  \state{Massachusetts}
  \country{USA}
  \postcode{02138}
}
\email{bumchul.kwon@us.ibm.com}


\begin{abstract}
The recent prevalence of publicly accessible, large medical imaging datasets has led to a proliferation of artificial intelligence (AI) models for cardiovascular image classification and analysis. At the same time, the potentially significant impacts of these models have motivated the development of a range of explainable AI (XAI) methods that aim to explain model predictions given certain image inputs.
However, many of these methods are not developed or evaluated with domain experts, and explanations are not contextualized in terms of medical expertise or domain knowledge.
In this paper, we propose a novel framework and python library, MiMICRI, that provides domain-centered counterfactual explanations of cardiovascular image classification models.
MiMICRI helps users interactively select and replace segments of medical images that correspond to morphological structures. From the counterfactuals generated, users can then assess the influence of each segment on model predictions, and validate the model against known medical facts.
We evaluate this library with two medical experts.
Our evaluation demonstrates that a domain-centered XAI approach can enhance the interpretability of model explanations, and help experts reason about models in terms of relevant domain knowledge.
However, concerns were also surfaced about the clinical plausibility of the counterfactuals generated.
We conclude with a discussion on the generalizability and trustworthiness of the MiMICRI framework, as well as the implications of our findings on the development of domain-centered XAI methods for model interpretability in healthcare contexts.
\end{abstract}




\received{20 February 2007}
\received[revised]{12 March 2009}
\received[accepted]{5 June 2009}

\maketitle

\section{Introduction}
In the recent decade, research studies on cardiovascular imaging have grown significantly due to the prevalence of publicly accessible, large datasets that have been made available to artificial intelligence (AI) researchers~\cite{fonseca_cardiac_2011, sudlow_uk_2015, sermesant_applications_2021, van_assen_artificial_2023, rouzrokh_machine_2023}.
Coupled with advancements in AI and machine learning (ML) methods, increasingly sophisticated models have been developed to support diverse clinical tasks from image segmentation to patient risk assessment and classification~\cite{sermesant_applications_2021, van_assen_artificial_2023, rouzrokh_machine_2023}.
At the same time, the potentially significant legal and ethical consequences of medical AI classification models \cite{miller2019medical, challen2019artificial, wachter2017counterfactual} have led to concerns about their trustworthiness, transparency, and interpretability \cite{antoniadi2021current, yang2022unbox, vellido2020importance}.
To address these concerns, a range of explainable AI (XAI) methods, such as saliency maps~\cite{ragnarsdottir_interpretable_2022}, image perturbation~\cite{li_brain_2018}, and example-based approaches~\cite{uehara_prototype-based_2020}, have been developed to explain model predictions given certain image inputs~\cite{salih_explainable_2023, chen_explainable_2022, van_der_velden_explainable_2022}.

However, these methods often do not contextualize explanations in terms of relevant domain knowledge, thus limiting their interpretability and usefulness to domain experts.
Saliency maps, for example, often produce outputs similar to simple edge detection, which can lead to risks of confirmation bias \cite{adebayo2018sanity}.
More seriously, many XAI methods and feature attribution tools are not evaluated with domain experts \cite{sperrle2021survey}, and do not improve human interpretations of model behavior \cite{casper2023red, bilodeau2024impossibility}.
Taken together, these limitations have led some researchers to conclude that ``most work in explainable artificial intelligence uses only the researchers' intuition of what constitutes a `good' explanation'' \cite{miller2019explanation}, highlighting the need for contextualized XAI methods that are developed for and interpretable to domain experts.
In the healthcare domain, for example, a cardiologist who wants to understand and evaluate a model trained to classify hypertension in cardiac magnetic resonance images (MRIs) may want to probe how morphological features, such as the various heart chambers, influence model predictions.
They may also want to validate model predictions against known medical facts, such as testing how patient age increases the predicted likelihood of hypertension.

Prior works have found that XAI methods centering user context and domain relevant concepts more effectively support model explanations \cite{ehsan2019automated, kim2018interpretability, lai2023selective}.
In this paper, we build on these existing approaches to develop the MiMICRI (\textbf{M}orphological \textbf{MI}xing for \textbf{I}nteractive \textbf{C}ounterfactual \textbf{R}ecombined \textbf{I}mages) framework for domain-centered counterfactual explanation of cardiovascular image classification models.
Counterfactual explanation is an XAI technique that, for a given input instance and an AI model, identifies the minimal perturbations necessary for the outcome predicted by the AI model to change.
They help users reason about the causes behind certain outcomes, and what it would take to change that outcome.
It has been argued that counterfactuals make for effective explanations because they are intuitive and actionable \cite{cheng2020dece, gomez2020vice}, compliant with legal regulations such as the GDPR\footnote{General Data Protection Regulation} \cite{wachter2017counterfactual, vellido2020importance}, and are model-agnostic \cite{wachter2017counterfactual}.

MiMICRI relies on state-of-the-art segmentation algorithms to identify the domain-relevant morphological features in an image that can be perturbed to generate counterfactual images.
To explain why a particular target image has a certain predicted label, domain experts can first select morphological segments to mask and replace with corresponding segments from other images, creating recombined instances.
The same predictive model is then used to generate new labels for the recombined images.
If a recombined image has a different label from the original target image, it is then a counterfactual of that target image.
Since the framework generates recombined images by replacing \textit{only} selected morphological segments, this allows users to attribute changes in model predictions to the replaced segments, thus providing an explanation of the model based on domain-relevant morphological image features.

To explore the effectiveness of our proposed MiMICRI framework, we implemented components of the framework as a Python package of the same name.
We then worked with two healthcare experts to evaluate the interpretability and effectiveness of the counterfactual explanations generated.
Overall, experts found that MiMICRI was helpful for reasoning about the relative influence of morphological features on model predictions, validating the model in context of known medical facts, and comparing between patient subgroups.
However, they also raised potential concerns around the clinical plausibility of the recombined images due to the structural and physiological interdependence of image segments.
In section \ref{discussion}, we discuss the implications of these concerns on the development of domain-centered XAI methods for model interpretability in healthcare domains.

To summarize, the main contributions of our work are the following: i) the MiMICRI visualization framework for counterfactual explanation and inspection of cardiovascular image classification models, ii) the MiMICRI Python visualization package, which provides an implementation of the proposed framework, iii) findings from an evaluation of MiMICRI with two experts, and iv) a discussion of the generalizability, trustworthiness, and implications of our findings for developing domain-centered XAI methods that enhance model interpretability in healthcare domains.

\section{Related Work}

This section examines recent advancements made in the fields of AI, explainable AI, and visual analytics applied to medical imaging data, with a focus on cardiovascular imaging.

\subsection{Cardiovascular image analysis using AI/ML/DL Models}

Recently, the number of research studies on cardiovascular imaging has grown significantly due to the prevalence of publicly accessible, large datasets and the rise of artificial intelligence (AI) methods.
In addition, the rapid developments in the field of computer vision have made it possible to apply neural networks trained with medical imaging data to a variety of tasks such as automated segmentation of cardiac structure, volumetric estimation, disease diagnosis, and outcome prediction (death or cardiac events). 
Various model architectures, ranging from convolutional neural networks to transformers, have been adapted for the purpose of predicting clinical outcomes based on information extracted from medical imaging datasets.
There are various types of medical images that are being used for such purposes, such as Cardiovascular Magnetic Resonance (CMR), echocardiography (ultrasound), computed tomography (CT), and nuclear imaging.
Active research in this field has become feasible, thanks in part to the open accessibility of imaging data and their corresponding clinical measurements being made available to AI researchers (e.g. UK Biobank~\cite{fonseca_cardiac_2011, sudlow_uk_2015}).
For more detail, readers may refer to the following reviews~\cite{sermesant_applications_2021, van_assen_artificial_2023, rouzrokh_machine_2023}.

\subsection{XAI approaches for medical image analysis}

Many XAI approaches have emerged within the field of computer vision and have been adapted for the analysis of medical imaging models. Previous surveys~\cite{salih_explainable_2023, chen_explainable_2022, van_der_velden_explainable_2022} offer comprehensive reviews of these XAI methods.
Of these, one of the most popular approaches for cardiac imaging analysis is the use of saliency maps (pixel-attribution maps). These model-based approaches visualize attention by employing various class activation mapping methods, and applying a heatmap to highlight the pixels in an input image that contribute most to the outcome.
For instance, Grad-CAM has been employed to identify the areas in echocardiograms of newborns that contribute the most to pulmonary hypertension~\cite{ragnarsdottir_interpretable_2022}.
Perturbation-based approaches, which are model-agnostic, explain predictions by replacing portions of an input image and displaying the predicted outcome of the altered image.
They have been employed to identify critical brain features contributing to autism spectrum disorders~\cite{li_brain_2018}.
In contrast, example-based approaches explain predictions by presenting a similar sample from the training data.
Still other researchers have proposed generating image patches from the training data and displaying prototypical examples that resemble a given image~\cite{uehara_prototype-based_2020}.

While these automated approaches can provide reasonable explanations, they may fall short when it comes to explaining why a specific instance is predicted as a certain class by the underlying model.
Pixel-attribution methods, such as saliency maps, assign importance to individual pixels, which may not be readily interpretable to clinicians.
Furthermore, these explanations have been found to resemble simple edge-detection \cite{adebayo2018sanity}, and do not pass benchmark tests for interpretability \cite{casper2023red}.
Feature perturbations have shown some promising results, but it can be challenging to automatically perturb images in a biologically plausible manner.
Similarly, example-based approaches may not elucidate how the model makes decisions based on the entire image, as they only display similar samples from prototype patches.

In response to these challenges, prior studies have shown that when working with tabular data, counterfactual explanations performed better than saliency maps and reduced over-reliance on ``wrong'' AI outputs during clinical decision making \cite{lee2023understanding}.
This suggests the potential for counterfactual image generation to also serve as an effective approach for explaining image classification models in the same context.
Many studies have proposed semantic counterfactual image generation approaches to explain AI models for general (i.e. non-medical) applications \cite{zemni2023octet, jacob2022steex, vandenhende2022making}.
These frameworks extract semantic regions from a target image and replace a subset of these regions to generate a counterfactual of the query image.
However, users often have no control over the image regions replaced, which may not always be relevant or appropriate for a particular domain application.
Additionally, in all frameworks, only a single counterfactual is produced to explain a target image.
This can limit the effectiveness of the frameworks since having multiple counterfactuals with slightly different perturbations provides a better explanation than a single sample \cite{wachter2017counterfactual, cheng2020dece}.
Furthermore, many existing counterfactual image generation techniques were also not developed for medical domains where there is a need to match and compare patients based on demographic characteristics \cite{xie_chexplain_2020}.
In these contexts, users may require more control over the source of the replacement regions in order to ensure that the characteristics of the source and target images/patients are sufficiently similar.

While there exists an approach, GANterfactual \cite{mertes2022ganterfactual}, which applies a similar segments-based counterfactual generation framework to medical images, GANterfactual automatically computes image segments using the SLIC algorithm \cite{schallner2020effect}, resulting in unlabeled segments that do not necessarily map to high-level features (e.g. morphological features).
In this paper, we build on existing frameworks to develop a domain-centered counterfactual generation approach for cardiovascular image classification models that provides interpretable and domain relevant explanations for medical experts.


\subsection{Visualization for XAI in (medical) image analysis}

To date, various visualization systems have been developed to explain AI trained on image data.
They help in identifying out-of-distribution samples \cite{chen_oodanalyzer_2020}, guiding clustering algorithms \cite{krueger_facetto_2020}, diagnosing and refining the layers of deep neural networks \cite{liu_analyzing_2018, liu_analyzing_2018-1, liu_towards_2017, wang_deepvid_2019, wang_dqnviz_2019}, searching for similar images \cite{cai_human-centered_2019}, identifying visual concepts~\cite{huang2023concept}, detecting and resolving biases~\cite{kwon2022dash}, and comparing the performance of generative adversarial nets \cite{wang_ganviz_2018}.
Of these visual analytics applications, some were developed specifically for the medical domain \cite{cai_human-centered_2019, xie_chexplain_2020, krueger_facetto_2020}.
Motivated by a need to explain clinical decision support systems and improve user acceptance of AI-assisted decision making, these prior studies collaborated with physicians and pathologists to design and build visualization systems that support expert-guided refinement of medical image search results \cite{cai_human-centered_2019} and image classification \cite{krueger_facetto_2020}, as well as the identification of key image features that influence AI image analysis \cite{xie_chexplain_2020}.
Findings from the CheXplain system \cite{xie_chexplain_2020}, in particular, demonstrated the effectiveness of comparative and contrastive examples for explaining model predictions.
However, CheXplain selects these contrastive examples from in-data set images instead of generating new counterfactual instances.
Furthermore, participants also mentioned that since physiological features may appear different for patients with different characteristics/demographics, there remained a need for tools that help them group patients and compare features by subgroup.

At the same time, a range of visual analytics systems have also been developed that broadly aims to provide counterfactual explanations of machine learning models across domains.
In these tools, users were able to guide the counterfactual generation process \cite{cheng2020dece}, analyze the importance of various features for model prediction \cite{zhang2023visual, gomez2020vice, wexler2019if}, and compare counterfactuals between user-defined subgroups \cite{cheng2020dece, zhang2023visual, wexler2019if, gomez2021advice}.
In evaluation studies, not only were these visual analytics systems found to provide effective explanations of AI models, the ability for subgroup comparison was also crucial for evaluating model fairness \cite{wexler2019if} and revealing bias \cite{cheng2020dece}.
However, the majority of existing systems were developed primarily for tabular data.
For example, some functions of the What-If Tool \cite{wexler2019if}, such as data perturbation and editing, could not be used on image data sets.
These limitations highlight the challenges of generating counterfactual explanations when working with image data.
The concept of a domain-relevant ``feature'' cannot be captured through pixel perturbations, which complicates the process of generating an interpretable and meaningful counterfactual through minimal edits.
In this paper, we address this challenge with MiMICRI, a domain-driven counterfactual explanation framework for cardiovascular image classification models.
We describe the details of this framework in the following section.






\begin{figure*}
  \includegraphics[width=\textwidth]{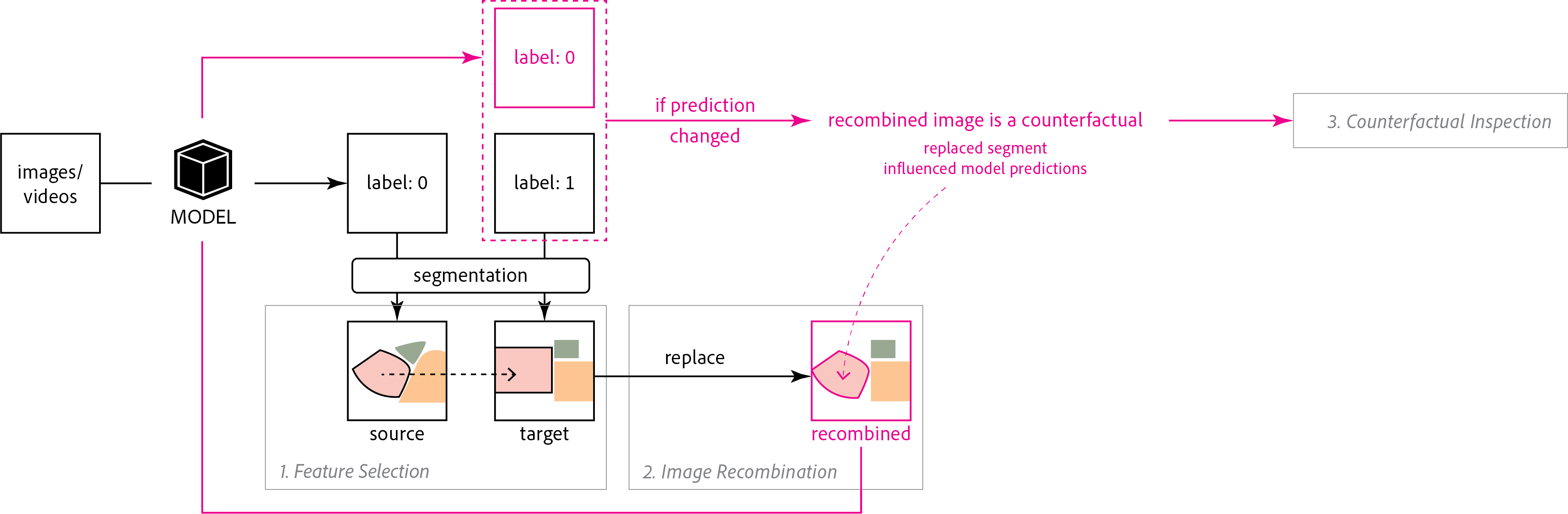}
  \caption{A high-level summary of the MiMICRI framework. To explain a classification MODEL, users can identify domain-relevant semantic image segments in each image in the data set, then replace segments in a target image with corresponding segments from a source image. This creates a recombined image. If the MODEL predicts that the recombined image has an alternate label to the target image, this recombined image is a counterfactual, and we can conclude that the replaced segment changed the MODEL prediction.}
  \Description{A diagram of the MiMICRI framework. On the far left, the diagram has a box labeled images/videos. An arrow points from the images/videos box to an icon labeled MODEL. An arrow extends to the right of MODEL to point at two boxes, labeled 0 and 1. Both the 0 and 1 boxes point to some text that reads ``semantic segmentation''. Below the text are two boxes with colored shapes. To the right of the boxes with colored shapes is a third box with colored shapes. This box has a label ``recombined''. The recombined box has an arrow pointing back to the MODEL icon, which then points to a box labeled 0. To the right of this is an arrow that reads: if prediction changed, recombined image is counterfactual.}
  \label{fig:mimicri_summary_simple}
\end{figure*}

\section{The M\MakeLowercase{i}MICRI Framework} \label{framework}


The MiMICRI framework (Fig. \ref{fig:mimicri_summary_simple}) is developed to explain decisions made by cardiac image classification models using counterfactual explanations grounded in relevant context and domain knowledge.
Consider an AI model trained to predict the likelihood of a patient having hypertension from their cardiac MRI data. Domain experts -- such as data scientists and healthcare providers -- may want to understand how the model is making predictions based on domain-relevant MRI image/video features.
These experts may also be interested in validating the model based on known medical knowledge.
For example, when doctors evaluate for hypertension in cardiac MRIs, they expect the LV myocardium to be thickened.
They may thus want to validate that changing this morphological feature influences model predictions in the expected direction (i.e. thinner LV myocardium decreases the likelihood of predicted hypertension and \textit{vice versa}).

\subsection{Counterfactual Generation Criteria}
To guide our initial development of MiMICRI, we identified three criteria from prior works that should be satisfied to produce a good counterfactual:

\begin{enumerate}[align=parleft, left=0pt..12pt]
    \item \textbf{Sparse/Minimal.} A counterfactual should make the minimum amount of changes needed to change the predicted label of an input image or video \cite{keane2021if, jacob2022steex, cheng2020dece, zemni2023octet, zhang2023visual, keane2020good, wachter2017counterfactual}. Users should be able to quickly identify the changes made, and reason about how they affected model predictions.
    \item \textbf{Plausible.} Also referred to as post-hoc validity, counterfactuals should be data instances that can realistically occur \cite{keane2021if, cheng2020dece, keane2020good}. For example, completely removing the left ventricle from an MRI will likely change the predicted model output, but this is not informative because such an MRI is unlikely to be obtained naturally from a patient.
    \item \textbf{Meaningful.} Since counterfactuals are meant to explain model predictions, this criteria ensures that the counterfactuals generated are sufficiently human-interpretable and explanatory \cite{jacob2022steex}.
\end{enumerate}

While these criteria are best-practices that apply to counterfactual generation regardless of data modality, it is not always clear how they should be extended to images and videos in particular.
For instance, a counterfactual image that is \textit{sparse/minimal} might not satisfy the \textit{meaningful} criteria.
Consider adversarial attack algorithms (for example \cite{huang2017adversarial, dong2018boosting, finlayson2019adversarial}) that make minor perturbations to image pixels to change the model prediction.
While these approaches might meet the \textit{sparsity} criteria, the changes are often undetectable to the human eye, and the resulting images are rarely referred to as counterfactuals since they do not provide \textit{meaningful} explanations of the model \cite{wachter2017counterfactual}.

The MiMICRI framework is inspired by existing image data augmentation methods such as CutMix \cite{yun2019cutmix}, Mixup \cite{zhang2017mixup} and Cutout \cite{devries2017improved} that identify a bounding box in an image that can be masked and/or replaced with pixels from another image, thus creating new images from available data.
However, in these prior methods, the resulting outputs are not designed to appear realistic (or \textit{plausible}), often producing images with missing patches \cite{devries2017improved} or areas that are visibly recombined \cite{yun2019cutmix, zhang2017mixup}.
MiMICRI extends these prior approaches by using segmentation algorithms trained on expert labels to identify domain-relevant morphological features to mask and replace.
In cardiac MRI data, for example, the left ventricle (LV) cavity, LV myocardium, and right ventricle (RV) cavity may be the relevant features to segment and identify.

\begin{figure*}
  \includegraphics[width=\textwidth]{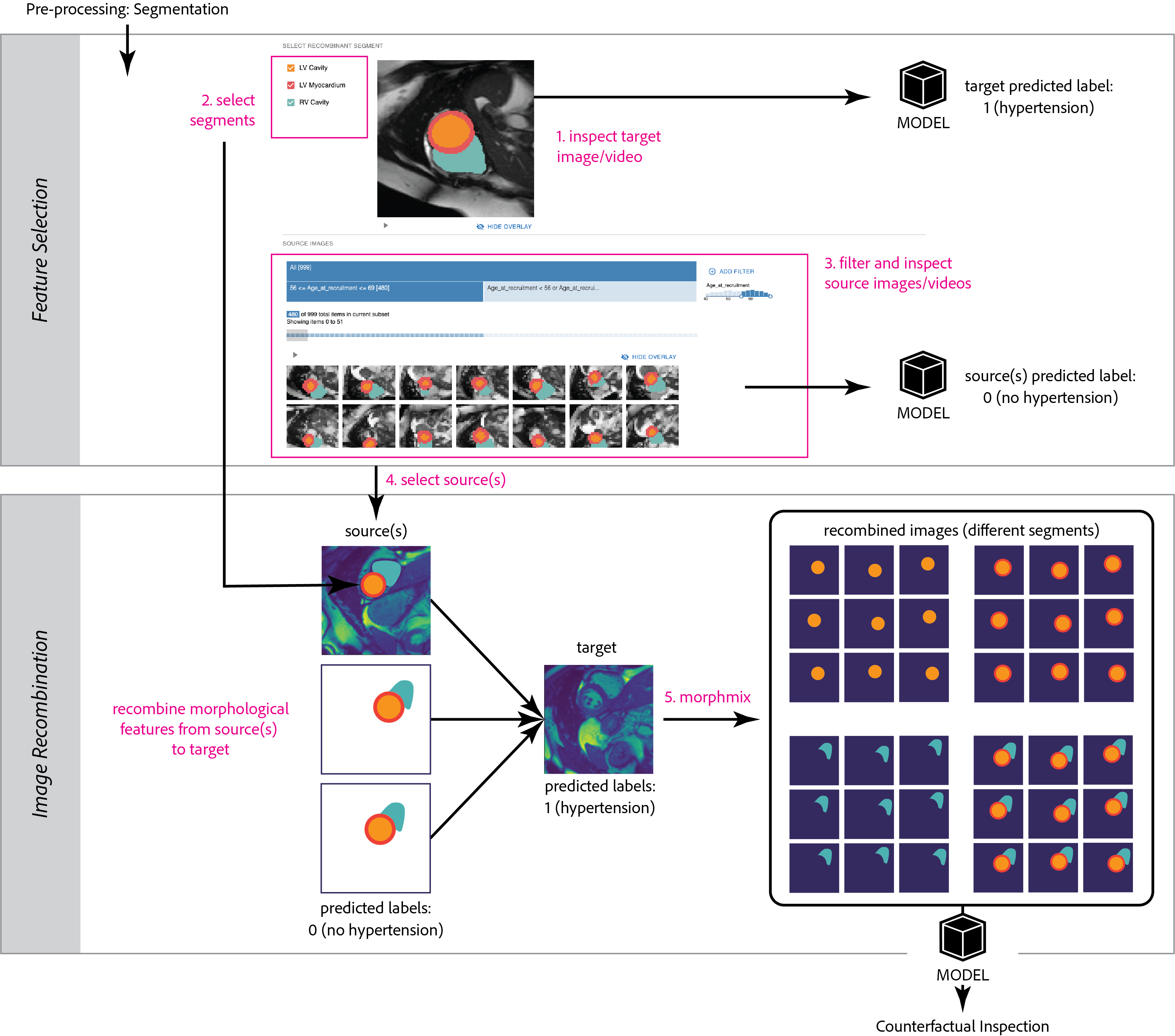}
  \caption{The detailed MiMICRI framework and corresponding visualization modules. \textit{Top:} Users interactively select source and target images or videos. They can also select combinations of segmented features to be replaced. Selected segmented areas (e.g. circular shapes in orange for LV Cavity) are overlaid on top of MRIs at their corresponding positions. \textit{Bottom:} Selected morphological segments from target(s) are masked and replaced with corresponding segments from source(s). We implemented the \textit{MorphMix} method to do this. New predicted labels are generated for the recombined images or videos. Users interactively inspect the model by viewing the counterfactuals generated for each replaced segment.}
  \Description{A diagram of the detailed MiMICRI framework. The diagram has three boxes laid out vertically. The top box is labeled ``MiMICRI selector'' and includes an annotated screenshot of a visual analytics tool. The middle box is labeled ``semantic image/video recombination'' and shows sets of source images, target images, and recombined images connected by arrows.}
  \label{fig:mimicri_summary}
\end{figure*}

To explain the predicted label for a particular target image, experts using MiMICRI can select and replace segments in the target image with corresponding segments from source images that have a different predicted label.
This replacement process generates new recombined images.
If a recombined image is also predicted to have a different label from the original target image, it is then a counterfactual of the target image.
Since only relevant morphological segments were replaced, and replacement segments were sampled from real (in-data) patient MRIs, this approach ensures that the counterfactual is \textit{plausible} and \textit{meaningful}.
Furthermore, since the rest of the original target image remains unaltered, the changes are also \textit{sparse/minimal}, allowing users to attribute any differences in model predictions to the replaced segments.
The MiMICRI framework has four main components: \textit{Image Segmentation (Pre-processing)}, \textit{Feature Selection}, \textit{Image Recombination}, and \textit{Counterfactual Inspection} (Fig. \ref{fig:mimicri_summary}).
We describe each component in detail in the following sections.
Our example uses MiMICRI to explain a model trained to predict hypertension likelihoods from cardiac MRI data.

\subsection{Pre-processing: Segmentation}
MiMICRI assumes that all images/videos in the data set have been segmented to identify the high-level semantic visual features relevant to the user and the domain.
While this pre-processing can be completed manually, it is also possible to train a segmentation algorithm based on expert labels.
In our example (Fig. \ref{fig:mimicri_summary}), three key morphological features are identified: the LV cavity, the LV myocardium, and the RV cavity.

\subsection{Feature Selection}
The framework starts with selecting the images/videos and cardiac segments for counterfactual generation.
Users can first select a target cardiac MRI with a certain model prediction (e.g. hypertension) to be explained (Fig. \ref{fig:mimicri_summary}, 1).
They can also select a set of source MRIs with the opposite predicted label (e.g. no hypertension) (Fig. \ref{fig:mimicri_summary}, 3 and 4), and 
the combination of segments to replace (Fig. \ref{fig:mimicri_summary}, 2).

\subsection{Image Recombination}
Having selected the source image(s) and segment(s) to replace, MiMICRI next generates all possible recombined images (Fig. \ref{fig:mimicri_summary}, 5).
In this paper, we developed the \textit{MorphMix} method (see Section \ref{morphmix}) for segment replacement, however, other methods can be considered in future work, such as training a generative adversarial network to inpaint a masked area.

\subsection{Counterfactual Inspection}
Finally, after generating multiple recombined images, the same hypertension model can be used to predict the disease likelihood of the recombined images.
Images with a predicted label different from the original target image would be considered counterfactuals.
Since the recombinations are identical to the target in all places except for the replaced segments, we can conclude that any counterfactual predictions (a different label from the original target) must be attributed to the replaced segments.
Furthermore, if experts selected multiple different combinations of segments for recombination, they can also compare the relative influence of different morphological features on model prediction.

\section{M\MakeLowercase{i}MICRI} \label{mimicri}

To implement the steps of the \tool framework, we built a Python visualization package\footnote{https://github.com/IBM/mimicri} that includes a selector module and the \textit{MorphMix} method.
The selector module helps users interactively select source and target images, as well as segments to be replaced.
Using the selected images and segments, the \textit{MorphMix} method then generates recombined images by replacing selected morphological segments in the target images with corresponding segments in source images.
The recombined images can be fed into the same classification model to generate new predicted labels.
In the following sections, we detail the implementation of the MiMICRI package.
MiMICRI uses the IPyWidgets\footnote{https://ipywidgets.readthedocs.io} framework and is designed to work in JupyterLab.
Module front-ends are implemented in React\footnote{https://react.dev/}, D3 \cite{bostock2011d3} and WebGL\footnote{https://www.khronos.org/webgl/}.


\subsection{Data Set, Hypertension Predictive Model, and Image Segmentation} \label{data_model}

To demonstrate MiMICRI, we trained a video classification model to predict hypertension from input MRIs.
Each MRI is a video with 50 image frames.
We used a 3D ResNet~\cite{he2016deep} for the model architecture comprising 50 layers with alternating 3D convolutional and batch normalization layers following the guidelines from the original paper.
For the dataset, we randomly split 23,043 patients' cardiac MRI scans of UK Biobank~\cite{sudlow_uk_2015} into training (18,434) and test (4,609) datasets.
We downsampled and standardized the MRI scans to 50$\times$128$\times$128 (frames$\times$width$\times$height).
For training, we selected a batch size of 16 to balance memory usage and computational efficiency. To facilitate convergence, we used the Adam optimizer with an initial learning rate of 0.001, which can be adjusted during training using learning rate schedulers, such as a step decay or cosine annealing schedule.
After training for 100 epochs, the model achieved a performance of accuracy 0.87 and auroc 0.65.
We saved the weights of the trained model and used it for the demonstration and use case in the following sections.
To protect the privacy of patients' healthcare data, we do not describe personally identifiable information of patients. 
We also exclude any original MRI files in the code repository and in system screenshots.
We segment all cardiac MRIs using the state-of-the-art \textit{ukbb\_cardiac} library~\cite{bai2018automated, bai2018recurrent}.

\subsection{Feature Selection: Selector}

\begin{figure*}
  \includegraphics[width=\textwidth]{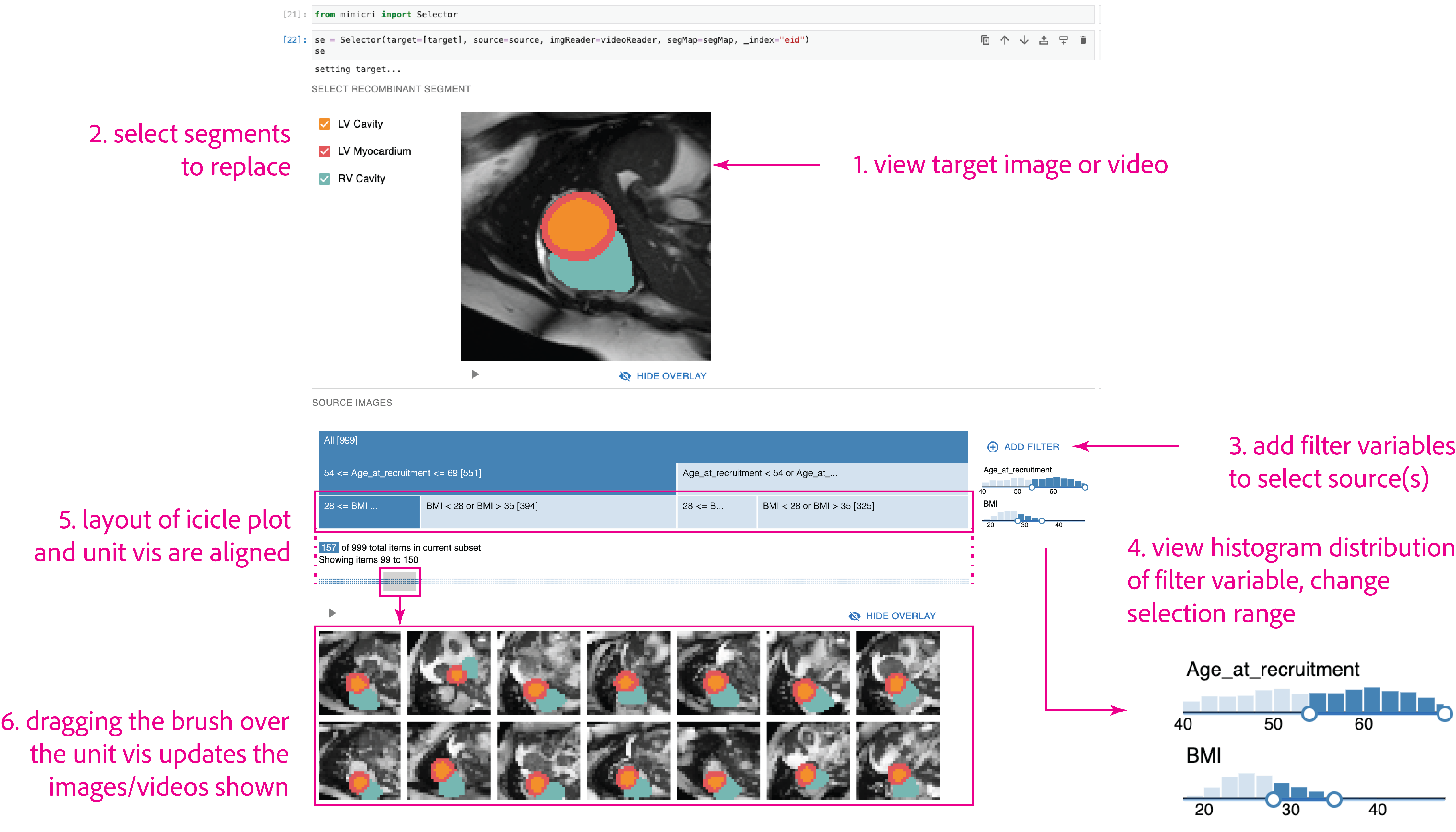}
  \caption{The MiMICRI selector module. In this module, users can 1) view a target image or video, 2) select segments to replace, 3) select source images by demographic by adding filter variables, 4) view and modify the range of selected values for each demographic filter, 5) view the selected subset in an icicle plot and corresponding unit visualization, and 6) view detailed source images or videos by dragging the brush over the unit visualization. In both the top and bottom panel, the visibility of the overlay can be toggled. If the files are videos, the videos can be paused. The panels can be resized. }
  \Description{This is an annotated screenshot of the MiMICRI selector module. The screenshot has two panels. The top panel shows a single video with checkboxes on the left. The bottom panel shows an icicle plot, a unit visualization, and a set of thumbnail size videos laid out in two rows and seven columns.}
  \label{fig:selector}
\end{figure*}

The selector visualization module (Fig. \ref{fig:selector}) is designed to help users systematically generate recombined images by selecting 1) the segments to replace, and 2) the source images to be used as replacements.
Before using the selector module, users should segment all files to identify domain-relevant morphological features.
In our example, we use the LV cavity, the LV myocardium, and the RV cavity segments.
Users can instantiate the module using the \verb|Selector()| function.
This function accepts as parameters the \verb|target| image to be explained, \verb|sources| with a different label from the target, an \verb|imgReader| function that converts all file paths to a \textit{numpy} array, and a \verb|segMap| object that includes the names of the segments identified.
On load, the selector module will display the target image in the top panel (Fig.~\ref{fig:selector}, 1).
The bottom panel is where users can filter, explore, and select source images for recombination.

\subsubsection{Select image segments for recombination.}
All segments identified during pre-processing are shown in overlay over the target and source images or videos.
Users can manually select the segments to be used in the subsequent MorphMix method (Fig. \ref{fig:selector}, 2).
This allows MiMICRI users to determine the image features that should be replaced based on domain knowledge and prior expertise.
Users can also toggle the visibility of the overlay for the target and source panels separately.
Once users have selected the segments they want to recombine, they can access the indices of the segments using the \verb|.segments| command in a subsequent JupyterLab notebook cell.

\subsubsection{Dynamically filter, explore and select source images for recombination.}
Using the drop-down menu in the bottom panel of the selector module, users can successively filter the source subset based on their demographic data (Fig. \ref{fig:selector}, 3) and view the subset of filtered images or videos.
The distribution of each variable is visualized as a histogram, and the range of selected values can be modified by dragging the range slider along the \textit{x}-axis of each histogram (Fig. \ref{fig:selector}, 4).
For example, if a target cardiac MRI is from an individual aged 65 years, we may want to select source MRIs from patients with a similar age range to control for any age-related differences.
Once a value range is selected, the icicle plot dynamically updates to reflect the cohort size after the new filters are applied (Fig. \ref{fig:selector}, 5).
Below the icicle plot is a unit visualization of all possible source images provided (Fig. \ref{fig:selector}, 5).
Selected source images are colored dark blue and left-aligned such that the unit visualization is visually consistent with the lowest layer of the icicle plot.
A gray rectangular brush can be dragged along the horizontal axis so that users can view the selected source images/videos in the tiled display below.
By design, $\sim$50 images or videos can be displayed at any time for rendering efficiency.
Once users are content with the selected sources, they can access the IDs of the selected items using the \verb|.subset| command in a subsequent JupyterLab cell.

\subsection{Image Recombination: MorphMix} \label{morphmix}

\begin{figure}
  \includegraphics[width=0.45\textwidth]{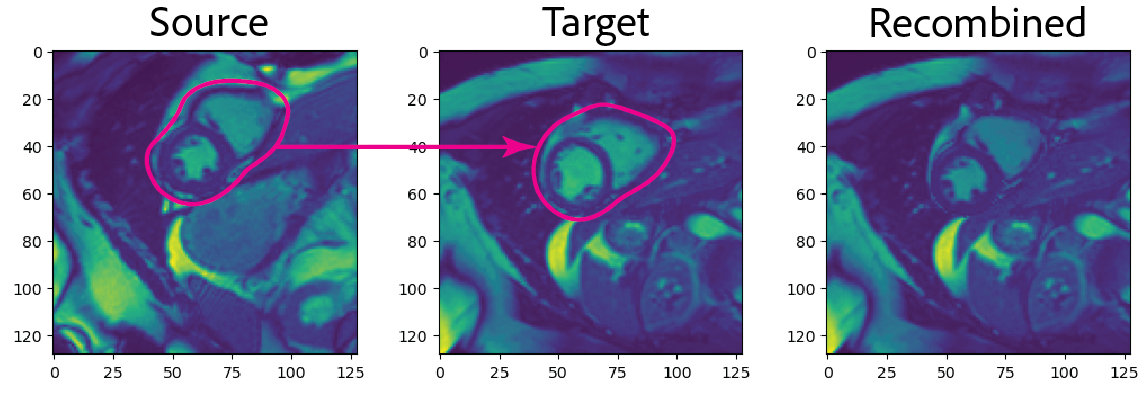}
  \caption{\textit{Left:} A single frame from a source cardiac MRI. \textit{Middle:} A single frame from a target cardiac MRI with all 3 cardiac segments to be masked. \textit{Right:} A recombined frame where pixels from the 3 corresponding cardiac segments in the source image were copied into the target image. Note how, with the exception of the replaced segments, the recombined image is identical to the target.}
  \Description{Three images laid out side by side. The image on left shows a cardiac MRI with part of the image outlined. The image in the middle shows a different cardiac MRI also with part of the image outlined. The image on the right shows a cardiac MRI that is a combination of the left and middle images. Nothing is outlined in this right-most image.}
  \label{fig:morphmix}
\end{figure}

After selecting the target, source(s), and segment(s) to replace, the \verb|Selector.morphmix()| function can be used to generate recombined images.
In the case of video data, each frame can be processed as a separate image.
The MorphMix method first masks the selected morphological segments of interest in an image or video -- that is to say, the pixels belonging to the segment(s) are removed (Fig. \ref{fig:morphmix}, \textit{Middle}).
More than one segment can be selected during this process, and different sets of segments can be tested combinatorially.
The method then replaces the masked target segment(s) with the corresponding segment(s) from the selected source image(s) (or video frame).
In our MorphMix implementation, we use a heuristic method to align the centroids of the segment(s) to be replaced. We then start at the centroids and use flood-fill to copy pixels from the source image into the target.
This outputs a recombined image with only the segment(s) of interest replaced.
The rest of the image pixels will be identical to the original target image (Fig. \ref{fig:morphmix}, \textit{Right}).


\captionsetup[table]{belowskip=5pt}

\begin{table*}
    \centering
    \begin{tabular}{lccc}
         segment(s) replaced&  counterfactuals (count)&  unchanged (count)& \% counterfactuals\\
         \hline
         LV cavity&  520&  2798& 0.157\\
         LV myocardium&  0&  3318& 0.000\\
         RV cavity&  496&  2822& 0.149\\
         LV cavity + LV myocardium&  639&  2679& 0.193\\
         LV myocardium + RV cavity&  762&  2556& 0.230\\
         LV cavity + RV cavity&  471&  2847& 0.142\\
         LV cavity + LV myocardium + RV cavity&  782&  2536& 0.236\\
    \end{tabular}
    \caption{Count and proportion of counterfactuals generated from 23226 recombined images. In general, replacing more segments influenced model predictions to a greater extent, resulting in more counterfactuals.}
    \label{tab:counterfactuals}
\end{table*}

\subsection{Counterfactual Inspection} \label{counterfactual_inspection}

Finally, new labels can be generated for all recombined images using the original predictive model to identify counterfactuals (i.e. recombined images with a different label from the original target image being explained).
In our example, we took the first 100 MRIs in the data set, of which 21 were predicted to have hypertension, and 79 were predicted to have no hypertension.
We ran \verb|Selector.morphmix()| using the hypertension group as sources and the no hypertension group as targets, replacing all possible combinations of the three identified cardiac segments.
We then repeated the run, switching the source and target groups.
In total, we generated 23226 recombined images (21 hypertension $\times$ 79 no hypertension $\times$ 2 runs $\times$ 7 segment combinations).
Using the same hypertension predictive model, we generated labels for all recombined images (Table \ref{tab:counterfactuals}).
The results correspond well to expected intuitions, where replacing more segments influenced model predictions to a greater extent, resulting in more counterfactuals generated.
This suggests that the hypertension predictive model in our example has, correctly, learned associations between cardiac features in an MRI and hypertension likelihoods.
However, there remain notable exceptions.
For example, unlike established medical cases, our results indicate that replacing the LV myocardium alone did not affect model predictions at all.
We discuss this further in expert evaluations (Section \ref{expert_evaluation}).

\section{Evaluation} \label{evaluation}

Since there are few methods for counterfactual generation in healthcare, and, to the best of our knowledge, none for cardiovascular imaging specifically, we evaluate the MiMICRI framework in two ways: first by inspecting the algorithmic segmentation outcomes of the recombined images generated, and secondly, through human validation in collaboration with two clinicians who have expertise viewing and assessing cardiac MRIs.

\subsection{Evaluation by Segmentation} \label{morphmix-segmentation}
Before presenting MiMICRI to experts, we first evaluated the generated recombined images by running the same \textit{ukbb\_cardiac} segmentation model on a random set of recombined images with different combinations of source, target, and replaced segments.
We then manually inspected the segmentation outputs.
As seen from Fig. \ref{fig:recombined_sample}, the segments identified corresponded well to their respective source and target images, providing initial validation that the framework produced recombined images of sufficient similarity to real MRIs for cardiac segments to be identified algorithmically.

\subsection{Expert Evaluation} \label{expert_evaluation}

We worked with two healthcare domain experts (2M) to evaluate the MiMICRI framework.
E1 is a pediatric cardiologist of 9 years.
His work involves performing clinical cardiovascular MRI (CMR) exams, deriving imaging-based biomarkers in congenital heart disease (CHD), as well as studying the use of wearable biosensors for CHD.
He uses AI models daily in CMR exams, specifically for image segmentation, but has yet to apply them to patient prognosis as such models are still under development.
E2 is a data scientist of 8 years in functional brain imaging, and who has recently transitioned to cardiac imaging.
His work mainly involves developing novel ML algorithms related to cardiac MRI acquisition and shape modelling of the heart, as well as supporting data extraction, transformation and loading processes for clinical outcomes research.
In prior work, E2 has used convolutional neural networks (VGG16 \cite{simonyan2014very}) and functional MRIs (a type of brain imaging) to measure differences between young and older adults during visual perception.

In a series of three meetings with the experts, we presented the end-to-end MiMICRI framework (Section \ref{mimicri}) using the hypertension prediction model for demonstration.
The meetings took place virtually, and lasted about an hour each.
Separately, we also provided a set of \textit{MorphMix} recombined cardiac MRIs to the experts for evaluation.
The recombined cardiac MRIs shared with the experts varied by source image, target image, and segments replaced.
In following sections, we discuss the expert feedback and concerns raised. 


\subsubsection{MiMICRI's framework and implementation provided more domain-relevant and interpretable explanations of model outputs than current methods.} \label{interpretability}
As part of his prior expertise, E2 has used existing explainability tools such as saliency maps \cite{simonyan2013deep} and Grad-CAM \cite{selvaraju2016grad, selvaraju2017grad} to inspect and evaluate the models he developed.
Compared to the previous methods, E2 found MiMICRI \textit{``less technically demanding''}, as it hides the technical details and allowed users to inspect model performance through the counterfactual explanations generated.
In contrast, tools like Grad-CAM expected users \textit{``to have certain knowledge about the structures of those models and the underlying machine learning frameworks''} (E2). 
Users of MiMICRI could also make domain-relevant manipulations of the counterfactual explanations that are relevant to the clinical scenarios, which is rarely offered by existing explainability tools.
More crucially for domain experts, these recombined images helped them interpret and validate the model in context of their domain knowledge.
When viewing Figure \ref{fig:morphmix}, for instance, E1 quickly reasoned that \textit{``one would hope that you're not looking at abdominal fat... if your AI model is only just looking at subcutaneous fat in the abdomen or in the chest wall, and then it's making a prediction based on that, it probably doesn't matter which part of the heart you put into the new image or not.''}
Taken together, both experts found that MiMICRI was more effective and interpretable than many existing approaches.


\subsubsection{Allowing users to select combinations of segments to replace across multiple source and target images supported more nuanced interpretation of model predictions.} \label{nuance}
During the evaluation, experts were particularly interested in the possibility of replacing specific segments of cardiac MRIs.
For example, E1 asked about combining \textit{``the small LV with the dilated RV''} or \textit{``a thick LV with a normal thickness RV''}, going on to explain that \textit{``those are probably going to be more physiologically explicable than taking the heart, the entire heart image, out of one person putting it in the other.''}
Additionally, E2 also mentioned that \textit{``an advantage of this toolbox is that it generates positive (counterfactuals) and negative (not counterfactuals) recombined images.''}
As such, though only some recombined images are counterfactuals, by selecting different combinations of source, target, and morphological segments, users can generate multiple recombined images that, when aggregated (as in Table \ref{tab:counterfactuals}), explain how each morphological feature affects model predictions.

\subsubsection{Filtering and creating subgroups from demographic data is important to physicians but less crucial to data scientists.} \label{filtering}
In clinical settings, AI models trained on imaging data may not always incorporate demographic information about the patient as input.
As such, E2, a data scientist, found that the filtering and subgrouping features in MiMICRI \textit{``might be unnecessary from a model development standpoint''} if the variable was not used during model training.
However, he also acknowledged that \textit{``demographic information of a patient is crucial in clinical settings.''}
This is confirmed by E1, a physician, who emphasized that when using MiMICRI, \textit{``you'd have to do that within people with similar other clinical characteristics so that you're not biasing your data set and then having [the model] predict off the clinical factors''}. 

In medical practice, BMI, weight, and abdominal fat, are all potential biological indicators of increased hypertension likelihoods.
For a physician using a hypertension predictive model, it is thus necessary to validate that the model provides \textit{``additional value''} (E1) beyond what is known about the patient.
These concerns apply to any specific clinical task or clinical prediction where \textit{``there's a whole bunch of confounders that come into play that you also have to account for in the image let alone in the clinical history''} (E1).
This also confirms prior works that found a need to compare patients with similar characteristics since physiological features may appear different between subgroups \cite{xie_chexplain_2020}.

\subsubsection{Results of the MorphMix method are not always clinically plausible since segments are structurally interdependent.} \label{plausibility}
While \textit{MorphMix} was able to generate recombined images where the replaced cardiac regions can be identified with high fidelity by our segmentation algorithm, E1 found that recombined cardiac MRIs were sometimes clinically implausible.
As he described: \textit{``the LV myocardium and blood pool (cavity) are completely interrelated because one bounds the other, so you can't change one without the other.''}
This likely also explains the unexpected result in Table \ref{tab:counterfactuals}, where replacing the LV myocardium alone did not generate counterfactuals because \textit{``the structures are interrelated, so you can't just pick the body of the LV out without changing the shape or features of the other objects''} (E1).

\begin{figure*}
  \includegraphics[width=\textwidth]{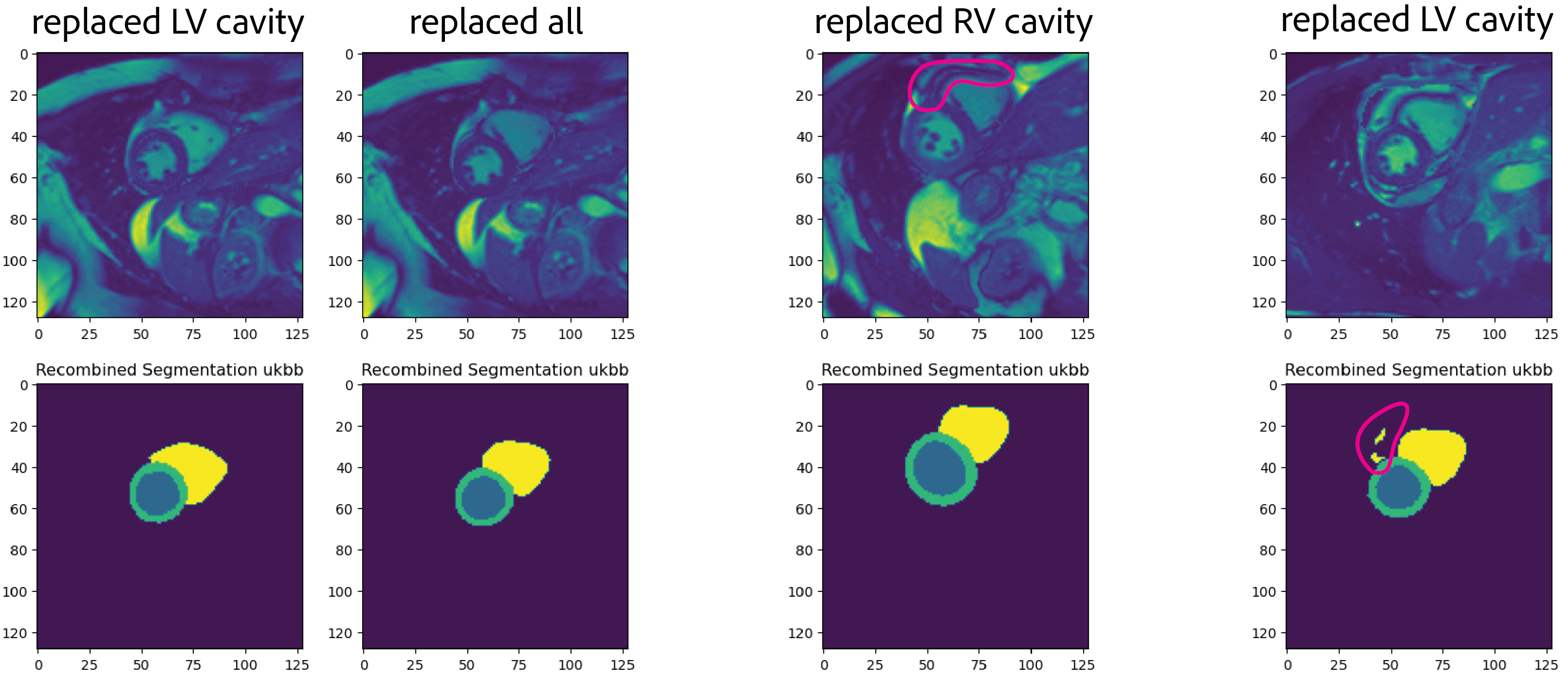}
  \caption{Expert feedback for four recombined images. Original source and target images were omitted for a compact layout. 
  \textit{Left:} Two recombined MRIs and corresponding segmentation that were acceptable to experts. \textit{Middle:} A recombined image with a double wall in the RV. \textit{``Though that may not affect segmentation, it would likely affect any whole-image analysis''} (E1). \textit{Right:} A particularly egregious example where physiological features were disordered and segmented regions contain artifacts.}
  \Description{Four pairs of cardiac MRI images and corresponding segmented image arranged horizontally. Each pair is laid out with the cardiac MRI above and segmented image below. Each segmented image shows three irregular shapes on a solid background. The third (middle) pair of images has a magenta region outlined on the cardiac MRI. The fourth (right-most) pair of images has a magenta region outlined on the segmented image.}
  \label{fig:feedback_small}
\end{figure*}

This highlights the challenge of recombining image segments when they are structurally interdependent.
Even in cases where the segmentation algorithm can accurately identify features in the recombined image, there may still be errors that are apparent to domain experts such as E1 (Fig. \ref{fig:feedback_small}, \textit{Middle}).
This suggests that generating recombined images that better meet the \textit{plausibility} criterion will require alteration to parts of the image surrounding the replaced segment.
However, this would come at a trade-off to requirements for \textit{sparse/minimal} changes, since image differences will no longer be limited to the segment being replaced.
In future work, a more in-depth study of different methods to mask and replace image segments may better balance the trade-offs between these criteria.

\section{Discussion} \label{discussion}

From expert evaluations, we found that the MiMICRI counterfactual explanations helped users reason about model predictions based on morphological structures and established medical knowledge (Sections \ref{interpretability} and \ref{nuance}).
These findings provide further support for frameworks proposed in prior works that center user context and domain relevant concepts to provide more effective model explanations \cite{ehsan2019automated, kim2018interpretability, lai2023selective}.
However, our expert evaluations also surfaced concerns about the generalizability, trustworthiness, and plausibility of our framework.
In this section, we discuss the implications of these concerns on how domain-centered XAI methods should be designed and developed.

\subsection{Generalizability} \label{generalizability}

Overall, while the experts found that the counterfactuals were useful, they also raised concerns about how well the MiMICRI framework applies to related data sets, such as other types of medical imaging.
To extend MiMICRI to other medical domains, generalist models, such as the Segment Anything Model (SAM) \cite{kirillov2023segment}, can be fine-tuned to specific types of medical imaging \cite{shi2023generalist} and diseases \cite{tandon2021retraining}.
Alternatively, recent advances in transformers for medical images \cite{ma2024segment}, active learning \cite{liu2023simpleclick}, and parameterized approaches \cite{lee2023scaling} can also be explored for segmentation.

At the same time, it must be highlighted that the \textit{MorphMix} method is most appropriate for organs that have \textit{``well defined anatomical structures that you can easily replace''} (E2), such as cardiac and skeletal structures.
It may not be as effective in generating recombined images in organs with complex contours, such as cortical foldings in the brain (E2).
Similarly, E1 also mentioned that \textit{``there are some organs or images that are more amenable to generating realistic/biologically plausible images, so... human oversight is needed to make the judgement [for when MiMICRI should be used].''}
This concern emphasizes the necessity of collaborating with domain experts to determine whether an XAI method would be applicable to a particular task and usage scenario.

\subsection{Model versus Explanation Trustworthiness} \label{trustworthiness}

More crucially, it is necessary to acknowledge that while the goal of many healthcare XAI tools, including MiMICRI, is to increase user trust in AI models \cite{cutillo2020machine, albahri2023systematic}, simply explaining a model does not necessarily result in greater trust. And nor should it.
In our approach, while the recombined counterfactuals helped medical experts reason about how the model made predictions, the counterfactuals were not always clinically plausible (Section \ref{plausibility}).
Or as E1 described, \textit{``you can manipulate the image but that's not what a real person looks like.''}
This implausibility led E1 to go on to comment that \textit{``I'm just worried about the implementation of MorphMix, not the underlying idea of it''}.
This demonstrates that while explanations can be useful, they may also be another source of error.
In our tool, some counterfactuals were valid while some were implausible (examples of both in Fig. \ref{fig:recombined_sample}), and we should take care that the recombined images are never presented as real MRIs.
Furthermore, we should also ensure that there is transparency in both model and explanation such that experts can evaluate the trustworthiness of both methods.

This distinction between model and explanation has been highlighted in prior works by Sperrle et al. \cite{sperrle2021survey} and Hohman et al. \cite{hohman2019gamut}. 
Hohman et al., in particular, found it concerning when participants were quick to rationalize explanations without questioning.
However, their evaluation was performed with randomly selected data scientists who were not necessarily domain (in their case, the housing market) experts.
In contrast, our work suggests that domain experts may be more cautious when evaluating XAI tools.
Such critical evaluations may be particularly important in medical domains where misplaced trust in erroneous predictions can result in serious adverse consequences \cite{larosa2018impacts, quinn2021trust, zhang2023ethics}.
Future work developing domain-centered XAI methods should thus look beyond just contextualizing explanations based on domain relevant information, but also ensure that both model and explanation are evaluated for trustworthiness with experts.


\subsection{Supplement, not Substitute, Expertise}

Finally, while the MiMICRI framework can be useful for explaining cardiovascular image classification models, the \textbf{explanations must be interpreted in context of domain knowledge, and should not substitute real-world clinical practice.}
In our hypertension example (Table \ref{tab:counterfactuals}), we found that replacing the LV myocardium alone generated no counterfactuals.
This contradicts known medical facts where LV myocardial thickness is one of the first things clinicians look at to see whether patients have the worst stage of hypertension (E1).
In this case, our results should not be taken as new medical claims.
Instead, they are more likely caused by the structural interdependence of cardiac segments (Section \ref{plausibility}), and should lead to a careful evaluation of the explanations presented.

Additionally, our process of developing and evaluating MiMICRI also revealed that \textbf{it is insufficient for XAI tools to rely solely on established best-practices or algorithmic evaluations.}
We had initially determined three criteria of ``good'' counterfactuals from prior work  (Section \ref{morphmix}), and later validated the recombined images by ensuring that constituent morphological features can be accurately identified using the same segmentation algorithm (Section \ref{morphmix-segmentation}).
However, concerns were still surfaced during the evaluation about the clinical plausibility of the counterfactuals (Section \ref{plausibility}).
Even in cases where the segmentation algorithm accurately re-identified cardiac segments, there were inconsistencies in the recombined MRI that were apparent to clinical experts (Fig. \ref{fig:feedback_small}, middle).
This highlights the gap in understanding between our definition of the \textit{plausibility} criterion and the expectations of domain experts.
It also reiterates findings from earlier work that XAI methods sometimes rely on ``researchers’ intuition of what constitutes a `good' explanation'' \cite{miller2019explanation} and may not meet the needs of intended users.
As such, for domain-driven XAI methods to truly enhance the interpretability and trustworthiness of AI models, best-practices and algorithmic evaluation are not enough -- they must be developed from an orientation that centers and builds on the knowledge and expectations of domain experts.

\subsection{Limitations}

While this work aims to provide domain-driven explanations of AI models in healthcare contexts, it is necessary to highlight that there are inherent risks associated with the use of AI in medical applications.
In radiology, for example, it has been found that \textit{``the occurrence of AI errors strongly influences treatment outcomes''} regardless of the experience of the radiologist or their familiarity with AI tools \cite{yu2024heterogeneity}.
While some studies have found that unlike other explainability techniques (such as saliency maps), counterfactuals effectively reduce over-reliance on `wrong' AI outputs during clinical decision making \cite{lee2023understanding}, there remains a need for us, as users and developers of XAI tools, to be cautious that the explanations do not lead to over-reliance on the AI.

Furthermore, as we better understand the uneven distribution of disease likelihoods in the population -- particularly for minoritized and underserved demographics -- it becomes increasingly important to validate model predictions for specific population subgroups to ensure more equitable outcomes.
Although our tool was designed for users to view data distributions and create subgroups via the histogram filtering feature, this approach is time consuming, and assumes that users know the demographics to inspect \textit{a priori}.
In future work, we plan to explore extensions to our tool that enhance subgroup analysis, such as algorithmic subgroup detection methods.
Finally, we also want to highlight that some groups, such as women and children, are underrepresented in AI algorithms and healthcare data sets in general.
This fundamental difference cannot be easily mitigated using our proposed framework, and would require more systematic changes in how we collect and curate data sets to ensure fair and equitable subgroup representation in clinical AI tools.



\section{Conclusion}
In this paper, we proposed the MiMICRI framework for conducting domain-driven counterfactual image analysis to understand cardiovascular image classification models.
This involves replacing high-level image features from one image into another to probe the relative importance of various morphological features for model predictions.
We implemented the framework as a Python visualization package, then evaluated its effectiveness with two medical experts.
Our findings highlighted the benefits of a domain-driven counterfactual explanation method, but also surfaced concerns about the generalizability and trustworthiness of our proposed framework.
We discussed these implications of our findings on the generalizability and trustworthiness of XAI methods, as well as the need to center and supplement domain expertise when developing tools for enhancing model interpretability in healthcare domains.

\begin{acks}
The authors would like to thank the Cleveland Clinic-IBM Discovery Accelerator partnership for their support.
We would also like to thank the anonymous reviewers for their thoughtful and detailed feedback on this paper, and members of the Georgia Tech Visualization Lab for their many helpful suggestions.
This work is supported in part by NIH/NHLBI K23HL150279, NSF IIS-1750474, IBM Research, and the IBM PhD Fellowship.
\end{acks}

\bibliographystyle{ACM-Reference-Format}
\bibliography{sample-base}

\appendix

\section{Evaluation by Segmentation}

\begin{figure*}[htb]
  \includegraphics[width=0.95\textwidth]{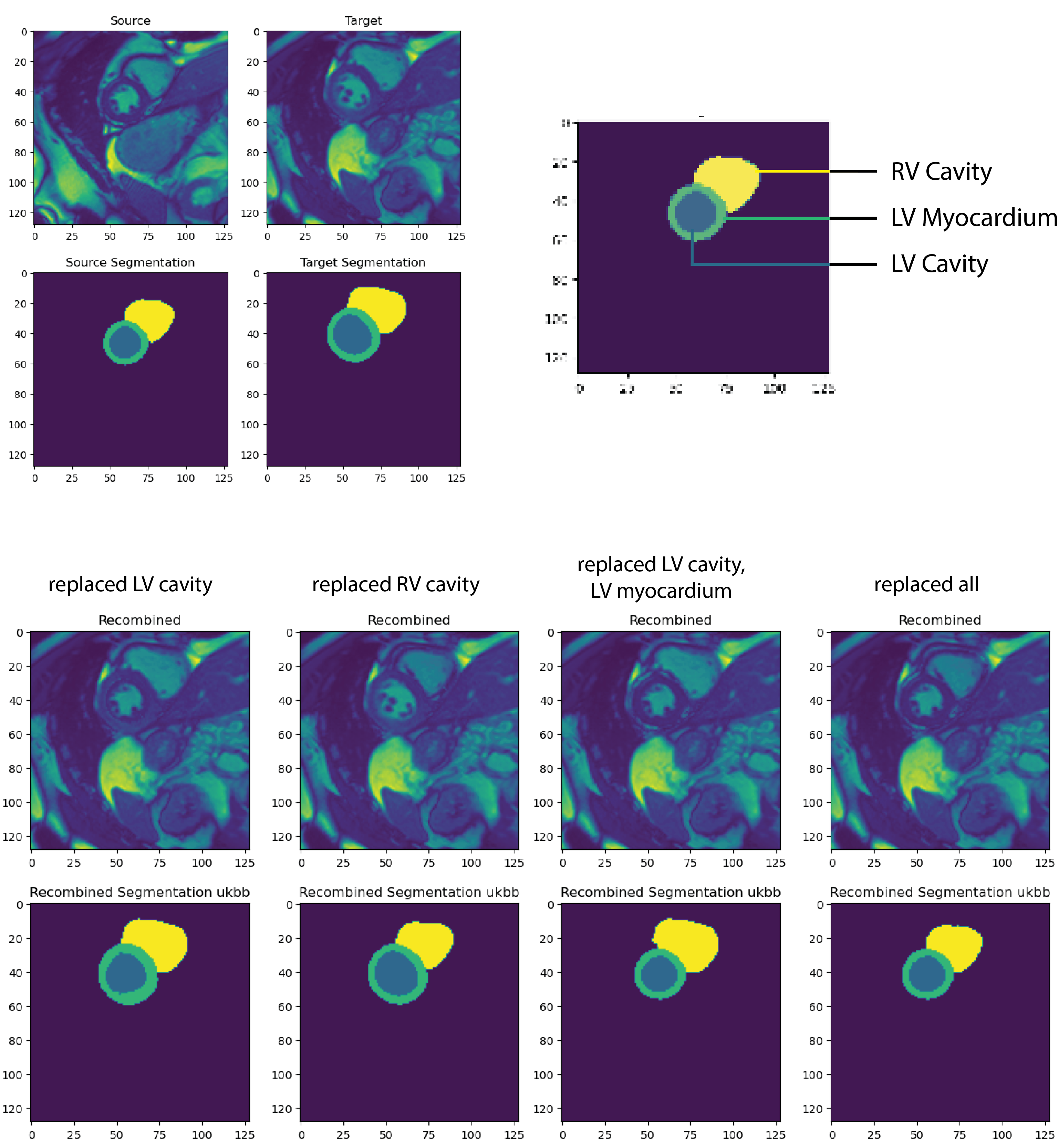}
  \caption{One pair of source and target MRIs and the resulting recombined images. As can be seen, the cardiac segments are re-identified by the segmentation algorithm with a high degree of fidelity. Note that due to space constraints, some infeasible combinations were excluded from this sample (e.g. replacing the LV myocardium alone since it is geometrically impossible to alter the LV myocardium without also altering the LV cavity enclosed within the myocardium).}
  \label{fig:recombined_sample}
\end{figure*}

To validate the quality of the recombined images, we selected random pairs of source and target images and replaced different combinations of cardiac segments for each pair.
We then used the \textit{ukbb\_cardiac} library to perform segmentation on the recombined images to ensure that the recombined morphological features can be identified (Section \ref{morphmix-segmentation}).
Figure \ref{fig:recombined_sample} shows one example of a pair of source and target images and the recombined images generated.

\end{document}